# Accumulated Polar Feature-based Deep Learning for Efficient and Lightweight Automatic Modulation Classification with Channel Compensation Mechanism

Chieh-Fang Teng, *Student Member, IEEE*, Ching-Yao Chou, *Student Member, IEEE*, Chun-Hsiang Chen, and An-Yeu (Andy) Wu, *Fellow, IEEE*

*Abstract*—In next-generation communications, massive machine-type communications (mMTC) induce severe burden on base stations. To address such an issue, automatic modulation classification (AMC) can help to reduce signaling overhead by blindly recognizing the modulation types without handshaking. Thus, it plays an important role in future intelligent modems. The emerging deep learning (DL) technique stores intelligence in the network, resulting in superior performance over traditional approaches. However, conventional DL-based approaches suffer from heavy training overhead, memory overhead, and computational complexity, which severely hinder practical applications for resource-limited scenarios, such as Vehicle-to-Everything (V2X) applications. Furthermore, the overhead of online retraining under time-varying fading channels has not been studied in the prior arts. In this work, an accumulated polar feature-based DL with a channel compensation mechanism is proposed to cope with the aforementioned issues. Firstly, the simulation results show that learning features from the polar domain with historical data information can approach near-optimal performance while reducing training overhead by 99.8 times. Secondly, the proposed neural network-based channel estimator (NN-CE) can learn the channel response and compensate for the distorted channel with 13% improvement. Moreover, in applying this lightweight NN-CE in a time-varying fading channel, two efficient mechanisms of online retraining are proposed, which can reduce transmission overhead and retraining overhead by 90% and 76%, respectively. Finally, the performance of the proposed approach is evaluated and compared with prior arts on a public dataset to demonstrate its great efficiency and lightness. The lightweight and efficient learning features of the proposed mechanism will be very attractive for future resource-constrained/aware IoT/V2X applications.

## I. INTRODUCTION

Fifth-generation (5G) cellular systems are expected to support over 50 billion devices by 2020, namely massive machine-type communications (MTC), which imposes a tremendous burden on base stations with the high cost of signaling overhead and energy consumption [1], [2].

To address the aforementioned challenges in enhancing the Quality of Service (QoS), software-defined radio (SDR), cognitive radio (CR), and systems with adaptive modulations have been studied extensively [3]-[8]. All of them tend to develop ***intelligent modems*** that are capable of doing spectrum sensing, self-adapting, and cooperation with neighbors, so as to fully utilize the available radio spectrum. Towards this goal, the mechanism of ***automatic modulation classification*** (AMC) is proposed as the intermediate step between signal detection and demodulation [9]-[12]. Based on the results of spectrum sensing, various modulation schemes are automatically classified to dynamically adjust transmission data rate and to meet the QoS requirement. At the receiver side, AMC is performed to blindly recognize the modulation types without prior knowledge of system parameters. Hence, with an effective AMC technology, the handshaking for exchanging *a priori* protocol information can be further reduced. It can achieve much lower signaling overhead and better transmission efficiency [10], which is especially useful for future IoT and resource-limited Vehicle-to-Everything (V2X) applications.

In the past decades, various research works have been proposed in the field of AMC. They can be generally classified into two classes: likelihood-based (LB) approach [9]-[17] and feature-based (FB) approach [9]-[12], [18]-[23]. The LB approach is based on maximization of the likelihood function with the assumption of the probability density function of an incoming signal. Though LB approach can provide optimal performance in the Bayesian sense, it requires perfect knowledge of the received signals and suffers from high computational complexity. On the other hand, FB approach, the decision is based on the extracted features of the received signals, such as higher-order statistics (HOS) and power spectral density. Compared with LB approach, FB approach is simpler for implementation with near-optimal performance.

In recent years, as machine learning (ML) techniques continually demonstrate significant breakthroughs in various fields, many researchers also exploit ML techniques as classifiers with extracted features, such as support vector machine, K-nearest neighbor, and genetic programming [24]-[26]. Without extracting hand-engineered features, deep learning (DL) can automatically learn the high-level features. It has received much attention due to its superior performance in recognition tasks with complex and deep architecture [27]-[36]. For example, one-dimensional convolutional neural network is exploited in [28], [30], [32], [36] to achieve promising performance with only raw IQ samples. Furthermore, [31] maps the received symbols to scatter points on the complex plane as the input for two-dimensional CNN with better performance as shown in Fig. 1(a).

This work is financially supported by the Ministry of Science and Technology of Taiwan under Grants MOST 105-2622-8-002-002, MOST 106-2221-E-002-204-MY3, and MediaTek Ph.D. Fellowship.

The authors are with the Graduate Institute of Electronics Engineering and Department of Electrical Engineering, National Taiwan University, Taipei, 10617, Taiwan (e-mail: {jeff, endpj, johnny}@access.ee.ntu.edu.tw; andywu@ntu.edu.tw).

TABLE I
COMPARISON OF AUTOMATIC MODULATION CLASSIFICATION FRAMEWORKS.

| Frameworks | Resource Consumption | | | Performance | |
|---|---|---|---|---|---|
| | Online Inference | Offline Training | Online Retraining | Under Ideal Channel | Under Channel Variation |
| Likelihood-based (LB) [13]-[17] | × | - | - | ◎ | × |
| Feature-based (FB) [18]-[23] | ◎ | - | - | △ | △ |
| Deep Learning-based [27]-[36] | × | × | × | ○ | △ |
| Proposed Accumulated Polar Feature with NN-CE | ○ | ○ | ○ | ○ | ○ |

◎: Optimum, ○: Good, △: Medium, ×: Bad

However, three critical issues of deep learning-based approaches are required to be addressed in practice:
1) *Heavy computational complexity*: DL-based approaches can achieve high recognition accuracy by taking advantage of the network to store intelligence learned from mass data. However, the complex and deep architecture induces severe computational complexity and memory overhead for online inference.
2) *Offline training overhead*: Another critical issue for DL-based approaches is the overhead of offline training phase, such as the training time and the necessity of mass training data, which consumes significant energy and severely hinders the deployment of DL-based approaches.
3) *Resource consumption of online retraining*: Only additive white Gaussian noise (AWGN) channel is assumed in the simulation of some prior works. It is good during the development stage, but not practical enough [29]-[31]. Besides, the real-world channels may vary over time which dramatically degrades the performance of DL-based approaches due to the mismatch between training data and inference data. Therefore, online retraining without intolerable resource consumption for channel adaptation is indispensable to ensure the model stays accurate.

In this paper, we aim to leverage the benefits of convolutional neural network for the problem of AMC. To address the three aforementioned issues, we present a novel approach for data transformation as shown in Fig. 1(b), and a neural network-based channel estimator (NN-CE) with an efficient online retraining mechanism to overcome the time-varying fading channel as shown in Fig. 1(c).[1] The main contributions of this paper are summarized as follows:
1) *Accumulated polar feature for deep architecture*: We propose *accumulated polar feature* by taking advantage of human's domain knowledge. The received signals are transformed into an easier classified domain, which helps to improve the classification accuracy to near-optimal region. Besides, with the help of data transformation, the model size, training time, and training data can be significantly reduced, when compared with [31]. Thus, we dramatically alleviate the computational complexity and lighten the training overhead by 99.8 times, which are more feasible for practical implementation and applications.
2) *Neural network-based channel estimator (NN-CE)*: We consider a fading channel with distorted amplitude and

[1] https://github.com/JieFangD/Automatic-Modulation-Classification

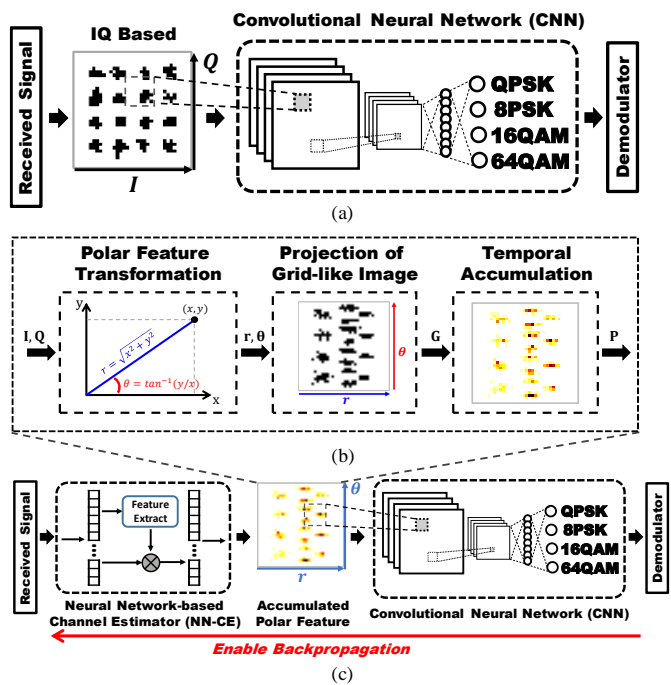

Fig. 1. Overview of (a) IQ based approach in [31], (b) proposed data transformation of accumulated polar feature, and (c) proposed accumulated polar feature-based deep learning with neural network-based channel estimator

phase shift which leads to severe performance degradation in related works. Then, we design a neural network-based channel estimator to recover the distorted channel with 13% improvement at SNR = 0dB.
3) *Efficient mechanisms for online retraining of NN-CE*: Considering the time-varying fading channel, we propose two efficient mechanisms for online retraining to remedy the degraded accuracy. With the benefit of NN-CE, we can reduce transmission overhead and retraining overhead by 90% and 76%, respectively.

We compare the computational complexity of the proposed approach with prior arts. It can achieve higher recognition accuracy while using much shorter inference time by 2,200 times compared with likelihood-based approach. Besides, the robustness under distorted channels can be effectively improved. Furthermore, the performance between different DL-based approaches is also evaluated on a public dataset for a fair comparison, which demonstrates the great efficiency and lightweight of our method. Therefore, our AMC mechanism is practical in resource-constrained applications such as V2X.

The comparison of related AMC frameworks is summarized in Table I. The rest of the paper is organized as follows. In Section II, the channel models and prior arts are introduced. The proposed deep learning architecture with data transformation is presented in Section III. In Section IV, the neural network-based channel estimator with the mechanisms of online retraining is introduced. Simulation results and analysis are conducted in Section V. The conclusions are finally drawn in Section VI.

## II. CHANNEL MODEL AND REVIEW OF PRIOR ARTS

### A. Notation

Throughout this paper, a normal-faced letter $a$ denotes a scalar, a bold-faced lowercase letter $\boldsymbol{a}$ denotes a vector, and a bold-faced uppercase letter $\mathbf{A}$ denotes a matrix. Other operations used in this paper are defined as follows:

- $\boldsymbol{a}[p]$ means the $p$-th element of $\boldsymbol{a}$.
- $\mathbf{A}[p,:]$, $\mathbf{A}[:,q]$, and $\mathbf{A}[p,q]$ denote the $p$-th row vector, $q$-th column vector, and $(p,q)$-th entry of $\mathbf{A}$, respectively.
- $\mathbf{A}^T$ denotes the transpose of $\mathbf{A}$.
- $|\cdot|$ denotes the element-wise absolute value or cardinality for a set.
- $[\mathbf{A}|\mathbf{B}]$ is the horizontal concatenation.

### B. Channel Model

We assume that baseband in-phase ($I$) and quadrature ($Q$) components of $\mathbf{y}[n]$ are extracted in a coherent, synchronous environment with single-tone signaling. In our approach, channel is treated as a flat fading in which frequency and phase offsets are added separately. The baseband sample of $\mathbf{y}[n]$ after matched filtering can be expressed as:

$$\mathbf{y}[n] = ae^{j(2\pi f_0 n + \theta)}\mathbf{s}[n] + \mathbf{g}[n], \quad (1)$$

where $a$ is an unknown amplitude factor, $f_0$ and $\theta$ are frequency offsets and unknown phase offsets, respectively, $\mathbf{s}[n]$ is transmitted symbol generated from one of selected modulations, and $\mathbf{g}[n]$ is the complex Gaussian noise. Therefore, the modulation classification task is to blindly identify these modulation categories merely from the $N$-sample received symbol vector $\mathbf{y} = [\mathbf{y}[0], \mathbf{y}[1], \cdots, \mathbf{y}[N-1]]^T$.

In this paper, we firstly consider four different modulation types for the preliminary evaluation and analysis of proposed approaches, including QPSK, 8PSK, 16QAM, and 64QAM, which are motivated by their difficulty in modulation classification literatures and widely used in many standards. Finally, we evaluate the performance on a public and difficult RadioML2018.01a dataset with 24 different modulation types as depicted in Section VI.

### C. Prior Art: Maximum Likelihood-based Approach [9]-[17]

Maximum likelihood-based (LB) approach can achieve optimal performance with Bayesian sense under ideal channel, such as additive white Gaussian channel, or perfect knowledge of channel parameters. Under AWGN channel, the probability of different modulation types can be easily derived as:

$$L(\mathbf{y}|m_i,\sigma) = \prod_{n=0}^{N-1}\sum_{k=1}^{M_i}\frac{1}{M_i}\frac{1}{\sqrt{2\pi}\sigma}e^{-\frac{|\mathbf{y}[n]-A_{i,k}|^2}{2\sigma^2}}, \quad (2)$$

$$m = \underset{m_i \in \mathcal{M}}{\operatorname{argmax}} L(\mathbf{y}|m_i, \sigma), \quad (3)$$

where $m_i$ is the $i$th modulation type in modulation pool $\mathcal{M}$, $\sigma$ is the variance of the signal-to-noise ratio, $A_{i,k}$ is the $k$th alphabet of the $i$th modulation, and $M_i = |A_i|$ denotes the alphabet size of the $i$th modulation. The classified result $m$ is based on the maximization of (2). Though LB methods have optimal performance, they suffer from high computational complexity for inference as well as high sensitivity to channel variation. Therefore, with the presence of fading channels, different likelihood ratio tests are proposed to overcome the problem of channel variation [9]-[17]. Among the different approaches, hybrid likelihood ratio test (HLRT) exploits a tradeoff between complexity and recognition accuracy. Then, (2) can be rewritten as:

$$L_{HLRT}(\mathbf{y}|m_i,\sigma)$$
$$= \max_{a_0 \in \alpha, \theta_0 \in \Theta}\prod_{n=0}^{N-1}\sum_{k=1}^{M_i}\frac{1}{M_i}\frac{1}{\sqrt{2\pi}\sigma}e^{-\frac{|\mathbf{y}[n]-a_0 e^{-j\theta_0}A_{i,k}|^2}{2\sigma^2}}, \quad (4)$$

where $\alpha$ and $\Theta$ are the set of unknown parameters of amplitude and phase, respectively. This method undergoes a long search process to find the unknown parameters which maximize the likelihood. The resolution of the set of the unknown parameters must be high enough to achieve better recognition accuracy. Therefore, the channel variation further increases the computational complexity of the LB methods. Finally, the analysis of complexity for different approaches will be compared and summarized in Section V.

### D. Prior Art: Deep Learning-based Approach [27]-[36]

Recently, deep learning (DL)-based approaches have been used to extract high-level features for AMC [27]-[36]. For example, one-dimensional convolutional neural network (CNN) is exploited in [28], [30], [32] and [36] to achieve promising performance with only raw IQ samples. Furthermore, [31] directly maps the received complex symbols to scatter points on the complex plane as the input for two-dimensional CNN which fully utilizes the characteristic of CNN with better performance as shown in Fig. 1(a). Although the method is simple and useful, learning from images in $I$-$Q$ plane loses the domain knowledge and known characteristics of the communication systems. Besides, it is apparent that the constellation of QPSK can be seen as a sub-picture of 8PSK, which is called nested modulations as shown in Fig. 2(a). Likewise, 16QAM also correlates with 64QAM. The shared constellation points in $I$-$Q$ based images result in high misclassification between them and consume more resources for offline training. Therefore, in our previous work [33], we have demonstrated that learning features from polar coordinates can improve recognition accuracy as shown in Fig. 2(b), which was also exploited in [34] and [35] for one-dimensional CNN and long short-term memory (LSTM), respectively.

Another practical problem is that only AWGN channel was considered in the simulation of some prior arts, which is unrealistic and results in inevitable degradation of recognition accuracy under the fading channel [29]-[31]. Besides, real-world channels may vary over time which leads to the mismatch between training data and inference data, thus requiring the online retraining to make up for the dramatically degraded performance of DL-based approaches. Therefore, the resource consumption of online retraining, including transmission overhead and retraining overhead, needs to be carefully considered and addressed.

## III. PROPOSED ACCUMULATED POLAR FEATURE-BASED DEEP LEARNING ARCHITECTURE

Before applying the input data to the concatenated CNN model, we transform the received signal to an easier classified domain, which significantly reduces both training overhead and model complexity. The process of data transformation for accumulated polar feature can be mainly decomposed into three parts, polar feature transformation, projection of grid-like image, and temporal accumulation, as depicted in Fig. 1(b).

### A. Construction of Grid-like Image in Polar Coordinates

*1) Polar Feature Transformation*: To deal with the aforementioned problem of resource consumption for offline training in $I$-$Q$ domain, we propose a polar feature transformation to map the complex symbols from Cartesian coordinates to polar coordinates. Learning features from $r$-$\theta$ domain encodes specific knowledge of the communication systems and makes the following convolutional neural network more robust to the fading channel. Taking spectral analysis for example, we certainly can input time-series data into learning process. The training result might lead to a Fast Fourier Transform (FFT) with large noise term. However, if we apply existing expert domain know-how, data can be transformed with FFT during data preprocessing prior to training. Therefore, the polar feature transformation can reduce the part of feature extraction in neural networks with better performance and lower training overhead as shown in Section V.

In this paper, by leveraging existed expert knowledge in communication, we can transform $I$-$Q$ domain into $r$-$\theta$ domain before learning. To construct the relation between $I$ and $Q$ components, we associate with polar coordinates which replaces the $I$-$Q$ axis with $r$-$\theta$ axis, as illustrated in Fig. 2. The process of transformation is summarized in Algorithm 1, where **I**, **Q** represent real part and imaginary part of received complex symbols, **r**, **θ** represent the transformed polar coordinates of radius and theta, and $N$ is the symbol length

---

**Algorithm 1:** Polar Feature Transformation

**Input I, Q**

**for** $n = 0 : N - 1$ **do**

  $\mathbf{r}[n] \leftarrow \sqrt{\mathbf{I}[n]^2 + \mathbf{Q}[n]^2}$

  $\mathbf{\theta}[n] \leftarrow arctan(\mathbf{Q}[n]/(\mathbf{I}[n]))$

**end for**

**return r, θ**

---

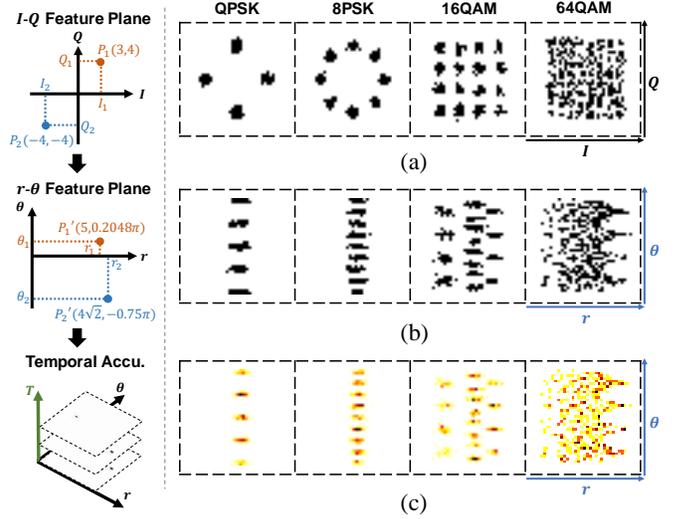

Fig. 2. Illustration of data transformation and constellation diagrams of four modulation categories with SNR=20dB: (a) $I$-$Q$ plane [31]; (b) $r$-$\theta$ plane [33]; (c) $r$-$\theta$ plane with temporal accumulation. (Note that darker color point represents higher values.)

---

**Algorithm 2:** Projection of Grid-like Image

**Input** $\mathbf{r}, \mathbf{\theta}, r_0, r_1, p_r, \theta_0, \theta_1, p_\theta$

**Initialize Image Matrix P** $= 0$

Compute grid interval of $r$ axis $\Delta g_r \leftarrow (r_1 - r_0)/p_r$

Compute grid interval of $\theta$ axis $\Delta g_\theta \leftarrow (\theta_1 - \theta_0)/p_\theta$

**for** $n = 0 : N - 1$ **do**

  Compute coordinate of $r$ axis $i \leftarrow \lfloor (\mathbf{r}[n] - r_0)/\Delta g_r \rfloor$

  Compute coordinate of $\theta$ axis $j \leftarrow \lfloor (\mathbf{\theta}[n] - \theta_0)/\Delta g_\theta \rfloor$

  **if** *polar_feature* **then**

    Compute pixel value of binary image $\mathbf{P}[i, j] \leftarrow 1$

  **else if** *accumulated_polar_feature* **then**

    Compute pixel value of grayscale image

    $\mathbf{P}[i, j] \leftarrow \mathbf{P}[i, j] + 1$

  **end if**

**end for**

**return P**

---

*2) Projection of Grid-like Image*: To utilize 2D-CNN for classification, we need to project the transformed symbols in $r$-$\theta$ axis to grid-like images. The method is similar to [31]. However, they set the image resolution to $227 \times 227$ as the input setting of AlexNet model [37]. Considering the prohibitively high computational complexity and the demand for low latency for communication, we design a specific convolutional model for the application of AMC and revise the image resolution from $227 \times 227$ to $36 \times 36$, which result in great performance and dramatically reduce the computational overhead compared with AlexNet model.

The process of projection can be summarized in Algorithm 2, where $r_0$, $r_1$ are the range of radius axis, $\theta_0$, $\theta_1$ are the range of theta axis, $p_r$, $p_\theta$ are the image resolution of two axes. Firstly, we compute the grid interval of two axes, $\Delta g_r$, $\Delta g_\theta$ with regard to the input settings. Secondly, each transformed symbol in $r$-$\theta$ axis is mapped to one point on grid-like image with $i, j$ representing the coordinate of $r$ axis and $\theta$ axis, respectively. Finally, the pixel value of grid-like image **P** is set to one, if there is any symbol mapped to the point. Therefore, we obtain the binary image, **P**, with the value set to either zero or one. The polar feature based images of selected modulation categories are depicted in Fig. 2(b), where the horizontal and vertical axes represent radius and theta, respectively.

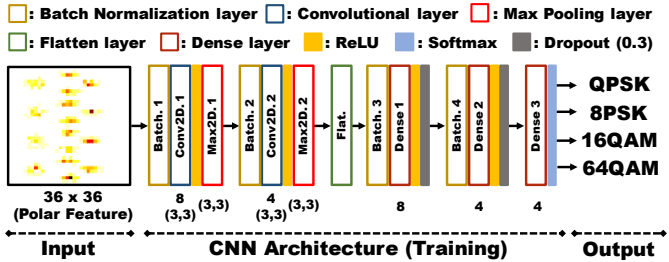

Fig. 3. The detailed architecture of the designed convolutional neural network.

*B. Temporal Accumulation*

To further increase the classification accuracy, we add another temporal axis to accumulate historical information of symbols in this dimension, which is not considered in our previous work [33]. The schematic is shown in Fig. 2(c). Therefore, the grid-like image is converted from binary to colorful, and the different colors represent the different probabilities of a symbol appearance at that point. Though the additive white Gaussian noise can be considered as a random process and degrade the classification accuracy, the noised symbol still has a higher probability to appear near the original point. With this approach, the higher probability of appearance has darker color as shown in Fig. 2(c), which makes the classification easier. The projection process of accumulated polar feature is also depicted in Algorithm 2.

Finally, **P** is the input data for the concatenated CNN model. In summary, with the polar feature transformation and temporal accumulation, we not only increase the classification accuracy but also reduce the size of neural network model and improve the convergence rate, which dramatically lower the offline training overhead of deep neural network as shown in Section V.

*C. Convolutional Neural Network (CNN)*

Deep learning (DL) is one of the fastest-growing fields with many breakthroughs recently. With different applications, there are many specifically customized architectures of neural networks. With that, convolutional neural network (CNN) is one of the widely adopt architectures in computer vision for object recognition and detection. A typical convolutional neural network architecture comprises of a number of convolutional layers for feature extraction, followed by fully connected layers (FC) for classification [37], [38].

The architecture of our CNN is constructed by two convolutional layers and three dense layers as illustrated in Fig. 3. The values below the convolutional layers represent the number and size of filters, and values below the dense layer represent the number of nodes. Before concatenating the next convolutional layer or dense layer, we add batch normalization layer to reduce training time and to avoid exploding or vanishing gradients [39]. More detailed descriptions of different layers can be found in [40]. The nonlinear activation function, Rectified Linear Unit (ReLU), among each layer is defined as:

$$\sigma_{ReLU}(x) = \max\{0, x\}. \quad (5)$$

Also, the activation function of softmax is to squash a *K*-dimensional input vector of arbitrary real values to *K*-dimensional vectors with each entry is in the range $(0,1)$ so that the summation of all entries is one. The softmax function is widely used in the output layer for classification to indicate the probability of each prediction. The function is defined as:

$$\sigma_{Softmax}(\mathbf{x}_j) = \frac{e^{\mathbf{x}_j}}{\sum_{k=0}^{K-1} e^{\mathbf{x}_k}}. \quad (6)$$

In addition, the regularization technique of "dropout", which avoids updating the weights of part nodes, is also utilized to reduce overfitting and force the nodes to be more independent than usual. Finally, the loss function $\mathcal{L}$, commonly categorical cross-entropy for classification tasks, is used to compute the error gradient between true labels **y** and predict labels **ŷ** and can be defined as:

$$\mathcal{L}(\mathbf{y}, \hat{\mathbf{y}}) = \frac{-1}{K} \sum_{k=0}^{K-1} \mathbf{y}_k \log(\hat{\mathbf{y}}_k) + (1 - \mathbf{y}_k) \log(1 - \hat{\mathbf{y}}_k). \quad (7)$$

For the training process, we use Adadelta optimizer [41] and the technique of early stopping [42] is utilized to avoid overfitting and ensure the convergence of neural network weights based on the performance on a validation set.

## IV. NEURAL NETWORK-BASED CHANNEL ESTIMATOR WITH ONLINE RETRAINING MECHANISM

In real-world wireless communications, there are many types of channel impairments, such as time dispersion, Doppler shifts, and inter-symbol interference. In this paper, we mainly focus on the impairments of the fading channel, namely power scaling and phase shift, which result in severe degradation of classification accuracy. Besides, the channel impairments are time-varying that result in a serious problem for DL-based approaches, due to the mismatch between training data and inference data. To address the two issues mentioned above, we design a neural network-based channel estimator to eliminate the channel effect and revise our deep architecture to make the online retraining feasible with great efficiency.

*A. Proposed Neural Network-based Channel Estimator (NN-CE)*

Considering the channel impulse response, the received constellation map must be distorted in amplitude and phase, which makes the recovery of the transmitted signal more challenging. Therefore, we design a neural network-based channel estimator to find out the inverse channel response and reduce the impact of power scaling and phase shift on the performance.

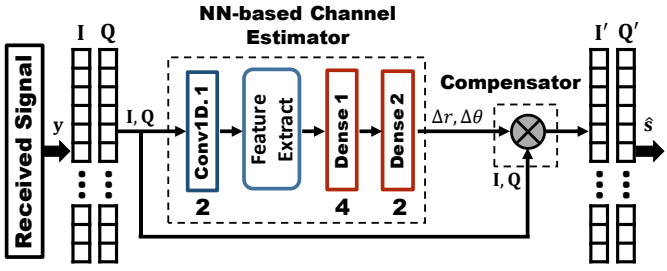

Fig. 4. The detailed architecture of neural network-based channel estimator (NN-CE) with two function-specific components: NN-based channel estimator and compensator.

In [43], the team of Google DeepMind proposed a spatial transformer network to learn an appropriate transformation for the input picture. The spirit of [43] is to design a localization network to estimate the transformation parameters. This is particularly useful for the application of image recognition where the input images may suffer from warping, rotation, and scaling. With the spatial transformer network, the model can be more spatially invariant to the input data. Additionally, there are also some works adopting this concept to design their own networks in order to overcome common channel impairments, such as radio transformer network in [44], [45] and channel compensation network in our previous work [33].

In this part, we design a neural network-based channel estimator to deal with the problem of fading channel as depicted in Fig. 4. Compared to [33], we further take advantage of the domain knowledge of communication systems to constrain the hypothesis class of the proposed network to reduce model size and avoid overfitting. It can be mainly divided into two function-specific components. The first one, NN-based channel estimator, is to estimate the compensated parameters $\Delta r$ and $\Delta \theta$. This can be achieved via some extracted features and the help of nonlinear transformation of neural networks. We firstly extract two features, average amplitude and standard deviation of the received signal, and feed them as the input for dense layers to transform the features to a higher level and estimate the compensated parameters. The second part, the compensator, takes the received signal as input and utilizes the estimated parameters to compensate for the distorted signal. The process can be expressed as:

$$[\mathbf{I}'|\mathbf{Q}'] = [\mathbf{I}|\mathbf{Q}] \times \begin{bmatrix} \Delta r\cos(\Delta\theta) & -\Delta r\sin(\Delta\theta) \\ \Delta r\sin(\Delta\theta) & \Delta r\cos(\Delta\theta) \end{bmatrix}. \quad (8)$$

The model is very tiny with only 44 parameters and suitable for online retraining to adapt to time-varying fading channel without massive resource consumption. And this will be introduced in Section IV.B.

For the offline training, we can train NN-CE first and concatenate it to the proposed deep architecture in Section III as shown in Fig. 1(c). For the training data, the input and desired outputs are the received signal with channel distortion and transmitted signal without any channel effect, respectively. Therefore, the purpose of NN-CE is to recover the distorted signal. The loss function is the mean absolute error and can be given by:

$$\mathcal{L}(\mathbf{s}_{golden}, \hat{\mathbf{s}}) = \frac{1}{N} \sum_{n=0}^{N-1} |\mathbf{s}_{golden}[n] - \hat{\mathbf{s}}[n]|, \quad (9)$$

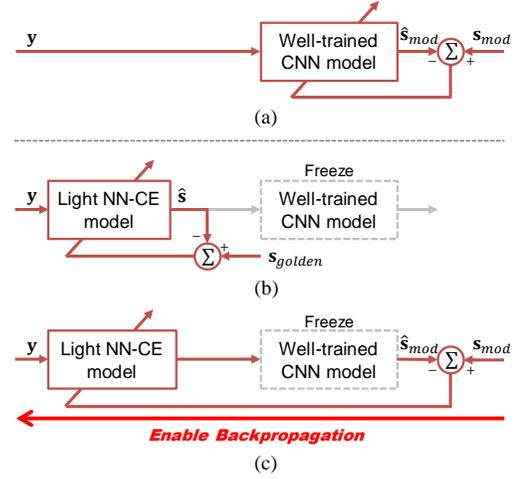

Fig. 5. Proposed mechanisms of online retraining: (a) retrain CNN without concatenation of NN-CE, (b) partially retrain NN-CE with $\mathbf{s}_{golden}$, and (c) partially retrain NN-CE with $\mathbf{s}_{mod}$ via end-to-end optimization.

where $\mathbf{s}_{golden}$ is the transmitted signal and $\hat{\mathbf{s}}$ is the compensated signal from NN-CE. For the time invariant channel, we can train a model in an offline environment to fit the channel optimally. However, for real-world applications, the channel may vary over time and degrade the recognition accuracy.

### B. Partial Retraining with Differentiable Transformation

To address the issue of time-varying channel, we need to retrain our model to maintain the high recognition accuracy. Besides, since online retraining is resource-limited, an efficient mechanism of retraining without intolerable resource consumption also needs to be carefully considered. Therefore, there are three different mechanisms of online retraining shown in Fig. 5 and we evaluate these mechanisms from two perspectives, namely in terms of transmission overhead and retraining overhead, which represent the occupied channel capacity for transmission of retraining data and the consumption of computing resource, respectively.

We first show that NN-CE can not only compensate for the distorted channel, but also make the online retraining more efficient. In Fig. 5(a), without the concatenation of light NN-CE model, we need to retrain the bigger CNN model with the labels of modulation types, which is also called $\mathbf{s}_{mod}$. This method requires more retraining data to avoid overfitting, thus inducing severe transmission overhead and retraining overhead, according to the uniform convergence with no free lunch theorem [46]. Therefore, to avoid the massive resource consumption, another two mechanisms with NN-CE model are proposed and shown in Fig. 5(b) and Fig. 5(c), respectively. In Fig. 5(b), in order to partially retrain NN-CE, we set training sequence $\mathbf{s}_{golden}$ as the desired output. This method is efficient and effective in making up for the degradation with little retraining overhead. However, it induces huge transmission overhead due to the transmission of longer training sequence $\mathbf{s}_{golden}$, which sacrifices the channel capacity and reduces the data throughput. Besides, the received signal $\mathbf{y}$ is only utilized

**Algorithm 3:** Projection of Differentiable Grid-like Image

**Input** $\mathbf{r}, \boldsymbol{\theta}, r_0, r_1, p_r, \theta_0, \theta_1, p_\theta$
**Initialize** Image Matrix $\mathbf{P} = 0$
Compute grid interval of $r$ axis $\Delta g_r \leftarrow (r_1 - r_0)/p_r$
Compute grid interval of $\theta$ axis $\Delta g_\theta \leftarrow (\theta_1 - \theta_0)/p_\theta$
**for** $i = 0 : p_r - 1$ **do**
   Compute grid value of $r$ axis $\mathbf{g}_r[i] \leftarrow r_0 + \Delta g_r \times i$
**end for**
**for** $j = 0 : p_\theta - 1$ **do**
   Compute grid value of $\theta$ axis $\mathbf{g}_\theta[j] \leftarrow \theta_0 + \Delta g_\theta \times j$
**end for**
**for** $n = 0 : N - 1$ **do**
   **for** $i = 0 : p_r - 1$ **do**
     **for** $j = 0 : p_\theta - 1$ **do**
       $\mathbf{P}[i,j] \leftarrow \mathbf{P}[i,j] + e^{\frac{-[(\mathbf{r}[n]-\mathbf{g}_r[i])^2 + (\boldsymbol{\theta}[n]-\mathbf{g}_\theta[j])^2]}{(2\sigma^2)}}$
     **end for**
   **end for**
**end for**
**return P**

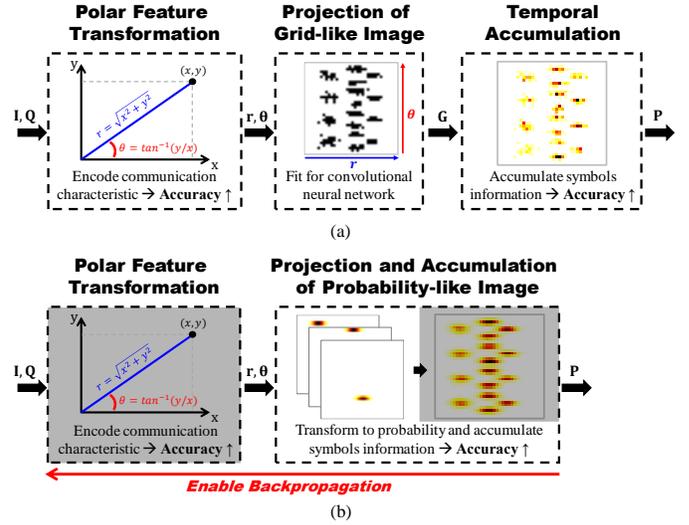

Fig. 6. Overview of (a) proposed data transformation of accumulated polar feature, and (b) revised transformation with enabled backpropagation.

in training phase without carrying any practical and effective information for communication.

To alleviate the transmission overhead, we propose another novel mechanism to partially retrain NN-CE without $\mathbf{s}_{golden}$ but with the label of modulation types $\mathbf{s}_{mod}$ as shown in Fig. 5(c). The main idea is to optimize the whole system in an end-to-end manner with partially updating the NN-CE model. With this approach, there are two issues that need to be solved: (i) connecting NN-CE with all transformation and the architecture of NN-CE to build up an integrated system, (ii) enabling the loss to backpropagate from CNN to NN-CE with partially updating. However, the transformation proposed in Section III is not differentiable, which prohibits the gradients of loss to propagate through the whole system. To resolve this issue, we need to revise the Algorithm 2 in Section III to make the whole process differentiable. The non-differentiable function is mainly caused by the absolute value function, which is used in computing the coordinate of $r$ axis and $\theta$ axis. Therefore, Algorithm 2 is revised to prevent the usage of absolute value function as summarized in Algorithm 3.

With the replacement of differentiable Gaussian distribution, we successfully enable the backpropagation from output to input and partially retrain NN-CE. The schematic of the revised process of transformation and patterns are shown in Fig. 6(b). Compared with Fig. 6(a), the patterns are smoother and can more accurately represent the occurrence probability of symbols on the picture.

Therefore, the shorter training sequence $\mathbf{s}_{mod}$ can dramatically reduce the transmission overhead with only a little additional retraining overhead for the end-to-end optimization. Besides, the received signal $\mathbf{y}$ can be used not only for the retraining of the model, but also for the transmission of practical information. For more detailed analyses of resource consumption, we compare the proposed mechanisms and evaluate the benefit of NN-CE in Section V.

TABLE II
SIMULATION SETTINGS

| Parameters | Notation | Values |
|---|---|---|
| Modulation Pool | $\mathcal{M}$ | QPSK, 8PSK, 16QAM, 64QAM |
| Training Data/Modulation | - | 5000 images |
| Testing Data/Modulation | - | 1000 images |
| Validation Ratio | - | 0.2 |
| Batch Size of Offline Training | - | 100 |
| Batch Size of Online Retraining | - | 10 |
| Image Resolution | $p_r, p_\theta$ | 36 |
| Range of Radius Axis | $r_0, r_1$ | [0, 3] |
| Range of Theta Axis | $\theta_0, \theta_1$ | [$-\pi/2$, $\pi/2$] |
| Amplitude Factor | $a$ | $U(0.2, 1)$ |
| Phase Offset | $\theta$ | $U(0, 2\pi)$ |
| Variation Degree | $\delta$ | 0.3, 0.5 |
| Symbol Length | $N$ | 1000 |
| Signal to Noise Ratio | $SNR$ | -4, -2, 0, 2, 4, 6, 8, 10, 12 |
| Training and Testing Environment | - | Deep learning library of Keras running on top of TensorFlow with i7-6700 CPU and NVIDIA GTX 1080 Ti GPU |

## V. SIMULATION RESULTS

In this section, we evaluate the performance of the proposed accumulated polar feature-based approach with prior arts through several simulations. We compare the classification accuracy and computational complexity among the different approaches.

Each simulation result is obtained from Monte Carlo trials over 1,000 independent channel realizations for each signal modulation. The classification decision is drawn from the modulation pool $\mathcal{M} = \{\text{QPSK}, \text{8PSK}, \text{16QAM}, \text{64QAM}\}$ as described in Section II.B. Therefore, the resulting classification accuracy is averaged over all 4,000 realizations. For the following experiments, if there is no specific statement, all parameters and environment settings are based on Table II.

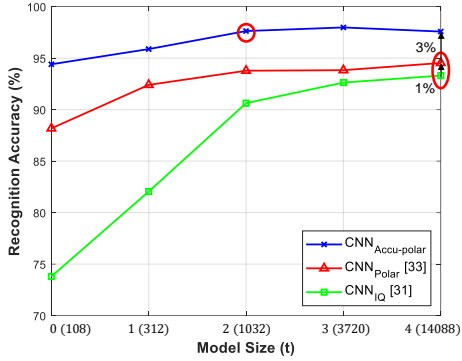

Fig. 7. Recognition accuracy against various model sizes when SNR = 8dB. Values beside model size represent the number of model parameters and the red circle indicates the selected model size for different approaches.

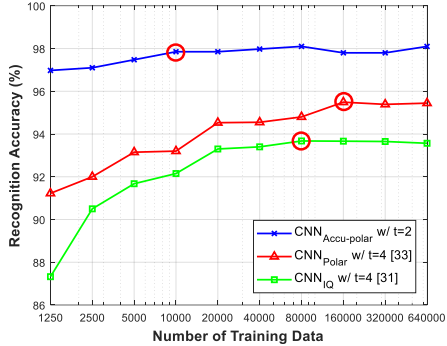

Fig. 8. Recognition accuracy against a various number of training data when SNR = 8dB. The red circle indicates the selected number of training data for different approaches.

### A. Reduction of Offline Training Overhead

Before comparing against prior arts, we firstly define a metric for the evaluation of offline training overhead:

$Offline\ Training\ Overhead$
$\triangleq (Model\ Size) \times (\#\ of\ Epoch) \times (\#\ of\ Training\ Data)$

With this defined metric, we compare the proposed accumulated polar feature-based approach with [31] and [33] via several experiments to evaluate the benefit of transformation descripted in Section III. Besides, for a fair comparison, the model architecture and image resolution for [31] and [33] are set as same as our approaches.

*1) Selection of Model Size*: Initially, we evaluate the performance under various model sizes with SNR = 8dB to depict the tradeoff between accuracy and offline training overhead and to decide the best model structure for us. Meanwhile, we evaluate the contributions of performance improvement from the techniques of polar feature transformation and temporal accumulation, respectively. Firstly, we fix the architecture of convolutional neural network as shown in Fig. 3. Besides, the number of filters for convolutional layer and nodes for dense layer are set to {*Conv1, Conv2,*

TABLE III
OFFLINE TRAINING OVERHEAD OF DIFFERENT DL-BASED APPROACHES

|  | IQ [31] | Polar [33] | Accu-polar |
|---|---|---|---|
| **Model Size ($t$) / Model Parameter** | 4 / 14,088 | 4 / 14,088 | 2 / 1,032 |
| **Number of Training Data** | 80,000 | 160,000 | 10,000 |
| **Accuracy** | 93.5 | 95.5 | 97.9 |
| **Time per Epoch (s)** | 4.40 | 8.07 | 1.08 |
| **Number of Epoch** | 15.33 | 14.67 | 16.78 |
| **Training Overhead** | 17,277,523K (99.8×) | 33,067,353K (190.9×) | 173,169K (1×) |

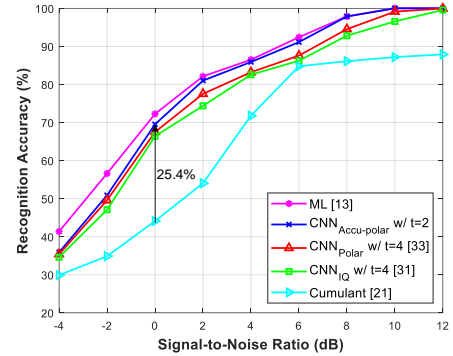

Fig. 9. Recognition accuracy against various SNR.

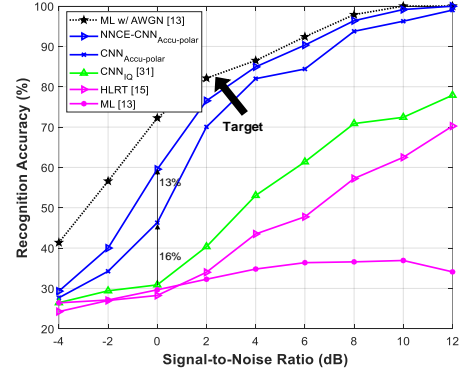

Fig. 10. Recognition accuracy against various SNR under fading channel.

*Dense1, Dense2, Dense3*}={ $2^{t+1}, 2^t, 2^{t+1}, 2^t, 4$ }, where $t$ varies from 0 to 4. In Fig. 7, we can observe that with the polar feature transformation, the signal is encoded to an easier classified domain with 1% improvement when $t = 4$. Next, the temporal accumulation can further enhance the performance by 3% with the accumulated historical information. For the following comparison, to highlight the tradeoff between performance and offline training overhead, the model size for IQ-based, polar-based, and accumulated polar-based are set to 4, 4, and 2, respectively. We can observe that our proposed transformation can effectively reduce model size and enhance performance.

*2) Number of Training Data*: Another important issue for DL-based approaches is the number of training data. Labeled training data for practical applications is very precious and rare, usually requiring great amounts of resources to obtain. In Fig. 8, we show that our proposed accumulated polar feature can converge with merely 10,000 training data. On the other hand, IQ-based and polar-based approaches require 80,000 and 160,000 training data for convergence, respectively. Furthermore, under a relatively low volume of training data, the proposed approach still achieves a significant performance improvement, which proves the importance of data transformation and the domain knowledge that are helpful in reducing both model size and required number of training data, by transforming data to an easier classified domain.

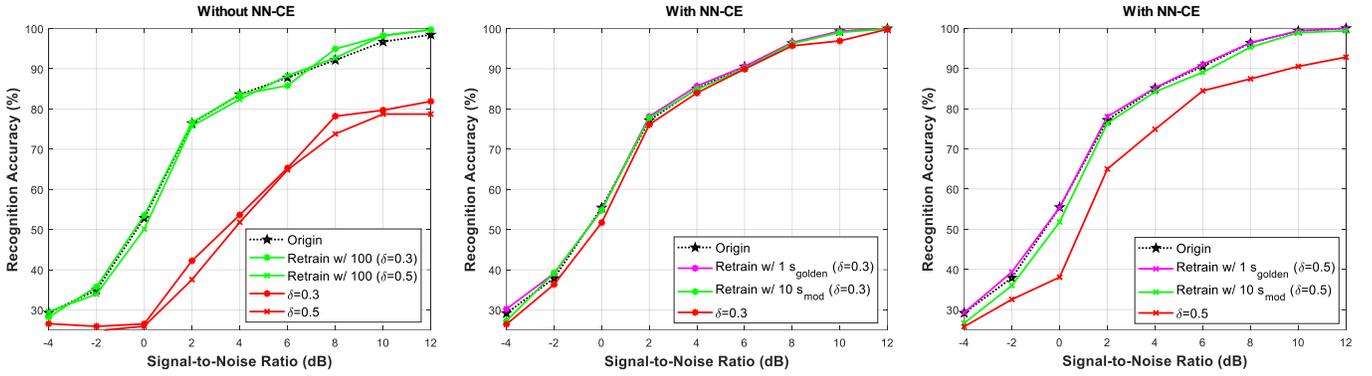

Fig. 11. Recognition accuracy of online retraining under time-varying fading channel: (a) online retraining without NN-CE, (b) proposed mechanisms of online retraining with $\delta = 0.3$, and (c) proposed mechanisms of online retraining with $\delta = 0.5$.

To sum up, a comprehensive comparison with evaluation metrics is shown in Table III. We can observe that the data transformation prior to training can significantly reduce the offline training overhead by about 100 times and 190 times compared to [31] and [33], respectively, having better efficiency and accuracy.

### B. Performance under Various Conditions

*1) Various SNR values under AWGN Channel*: With the chosen model size, we compare our proposed approach to prior arts under various SNR. In Fig. 9, we can observe that the LB method [13] can achieve optimal performance in Bayesian sense under AWGN channel. Therefore, we can treat [13] as the upper bound for the recognition accuracy. Simultaneously, we compare the performance to high-order cumulants (HOCs) with fourth order ($C_{40}, C_{42}$) and sixth order ($C_{63}$), which are detailed explained in [9]-[12], [18]-[23]. From the simulation results, our proposed method is far better than FB method [21] by 25.4% when SNR = 0dB, and is very close to the upper bound under high SNR conditions. Besides, our proposed accumulated polar feature is about 3% better than IQ-based approach [31] under the whole range of SNR values with a smaller model size. Due to the poor performance of FB, we do not compare it in the following experiments.

*2) Various SNR values under Fading Channel*: To make the system more practical, we consider the fading channel with a variance of received power and phase shift that is a common phenomenon in communication systems. The setup of variances $a$ and $\theta$ are shown in Table II, and other settings are kept the same as given in the previous experiment. Besides, we also evaluate the performance of the proposed approach with NN-CE, which can learn the inverse channel response and compensate for the distorted channel. The simulation results are shown in Fig. 10. When considering the fading channel, the performance degrades significantly, especially for the LB method which suffers from channel mismatch. Despite adopting HLRT method described in [15], it still experiences severe performance degradation under fading channels. As a tradeoff between the complexity and recognition accuracy, the spacing of unknown parameters of amplitude and phase for HLRT are set to 0.05 and 1, respectively. In contrast, our proposed accumulated polar feature outperforms all other methods by more than 16% under the most range of SNR, which demonstrates that our method is robust and tolerant to the channel distortion. Besides, the NN-CE can compensate for the channel distortion, which further improves the recognition accuracy by 13% when SNR = 0dB under a practical environment and approaches the upper bound of the ideal channel under high SNR conditions.

### C. Performance under Time-varying Fading Channel with Different Mechanisms of Online Retraining

Since channels may vary over time in real-world environments, we need online retraining to maintain the high recognition accuracy. NN-CE can not only compensate for the channel distortion, but also provide an efficient way for online retraining. In this part, we evaluate the proposed mechanisms of online retraining under different degrees of channel variation with varying sizes of retraining data. We first randomly generate an independent channel as the current channel according to the distribution listed in Table II and the model is well-trained under this setting. To model the effects of a time-varying channel, the amplitude factor and phase offset change according to the setting of variation degree, namely the parameters of fading channel will become $a' = a \times (1 \pm \delta)$ and $\theta' = \theta \times (1 \pm \delta)$ where the positive and negative signs are randomly decided. Therefore, the time-varying channel severely degrades the recognition accuracy due to the mismatch between training data and inference data. For the following reason, online retraining is required to make up for the degraded performance.

*1) Time-varying Fading Channel without NN-CE*: Firstly, we evaluate the proposed accumulated polar feature without the help of NN-CE, which means that the model is more sensitive to channel variation and require more retraining data to avoid overfitting. In Fig. 11(a), each point is the average of 4000 realizations under five different channels. We show that the performance degrades dramatically under 0.3 or 0.5 channel variation, and at least 100 data cases of each type of modulation are required to retrain the model back to the original performance. Though the required retraining data is less than training data, it still sacrifices the channel capacity for the transmission of retraining data and requires massive computing resource for the process of retraining.

TABLE V
NUMBER OF OPERATORS AND OPERATION TIME REQUIRED BY DIFFERENT CLASSIFIERS

| Channel | Classifier | Additions | Multiplications | Exponentials | Logarithms | Comparisons | Memory | Inference Time with CPU (s) |
|---|---|---|---|---|---|---|---|---|
| AWGN | ML [13] | $N(4\sum_{i=1}^{M}M_i + M)$ $\sim 10^5$ | $N(7\sum_{i=1}^{M}M_i + 4)$ $\sim 10^6$ | $N\sum_{i=1}^{M}M_i \sim 10^5$ | $NM \sim 10^3$ | $M = 4$ | $\sum_{i=1}^{M}M_i = 92$ | 0.00972 |
| | Cumulant [21] | $6N \sim 10^4$ | $16N \sim 10^4$ | 0 | 0 | $M = 4$ | $3M = 12$ | 0.00036 |
| | IQ [31] | $809p_rp_\theta + M + 2N$ $\sim 10^6$ | $809p_rp_\theta + M + 3N$ $\sim 10^6$ | $M = 4$ | 0 | $116p_rp_\theta + N$ $\sim 10^5$ | $14020 + 17M$ $= 14088$ | 0.00132 |
| | Accu-polar | $p_rp_\theta(4N + 107) + M$ $\sim 10^7$ | $p_rp_\theta(6N + 106) + M$ $\sim 10^7$ | $p_rp_\theta N + M$ $\sim 10^6$ | 0 | $29p_rp_\theta \sim 10^4$ | $1012 + 5M$ $= 1032$ | 0.00773 |
| Fading | HLRT [15] | $N_aN_\theta N(4\sum_{i=1}^{M}M_i + M)$ $\sim 10^9$ | $N_aN_\theta N(9\sum_{i=1}^{M}M_i)$ $\sim 10^9$ | $2N_aN_\theta N\sum_{i=1}^{M}M_i$ $\sim 10^8$ | $N_aN_\theta NM$ $\sim 10^6$ | $N_aN_\theta M$ $\sim 10^3$ | $\sum_{i=1}^{M}M_i = 92$ | 17.11982 |
| | Accu-polar with NN-CE | $p_rp_\theta(4N + 107) + M$ $\sim 10^7$ | $p_rp_\theta(6N + 106) + M$ $\sim 10^7$ | $p_rp_\theta N + M$ $\sim 10^6$ | 0 | $29p_rp_\theta \sim 10^4$ | $1056 + 5M$ $= 1076$ | 0.00775 |

*2) Time-varying Fading Channel with NN-CE*: With the benefits from NN-CE, two different mechanisms of online retraining, partial retraining NN-CE with $\mathbf{s}_{golden}$ and end-to-end optimization with $\mathbf{s}_{mod}$, have been detailed in Section IV.B. First, from Fig. 11(b) and Fig. 11(c), we can observe that the degradation is less compared to Fig. 11(a), which means that NN-CE can make our system more robust to channel variation. Second, with the label of training sequence $\mathbf{s}_{golden}$, we can effectively retrain our system back to original performance with only 1 data case for each type of modulation, consuming very little computing resources. However, the required long training sequence still induces severe transmission overhead. On the other hand, though the approach of end-to-end optimization requires 10 labeled data samples specifying the modulation types $\mathbf{s}_{mod}$ for each type of modulation, the transmitted data can be used not only for retraining the model, but also for the transmission of practical information, which dramatically reduces the transmission overhead with minimal additional retraining overhead.

### D. Analysis of Online Retraining

In this part, we further evaluate the resource consumption of different mechanisms of online retraining from two perspectives, transmission overhead and retraining overhead, which represent the occupied channel capacity for transmission of retraining data and consumption of computing resources, respectively. The two metrics can be defined as:

$Transmission\ Overhead \triangleq Occupied\ Channel\ Capacity$

$= (\#\ of\ Retraining\ Data) \times (Length\ per\ Data),$

and

$Retraining\ Overhead \triangleq Retraining\ Time.$

Table IV lists the transmission overhead and retraining overhead of three different mechanisms of online retraining as shown in Fig. 5. The required data length of $\mathbf{s}_{mod}$ is $\log_2 M$, representing the different types of modulations. On the other hand, $\mathbf{s}_{golden}$ must be the same length as the symbol length of $\mathbf{y}$ for the retraining of NN-CE. Furthermore, $\mathbf{y}$ also needs to be accounted as retraining data because it is only utilized in the training phase without any actual information.

TABLE IV
RESOURCE CONSUMPTION OF DIFFERENT MECHANISMS OF ONLINE RETRAINING

| | Retrain CNN w/o NN-CE | Partially Retrain NN-CE | |
|---|---|---|---|
| Label | $\mathbf{s}_{mod}$ | $\mathbf{s}_{golden}$ | $\mathbf{s}_{mod}$ |
| Length per Data (bit) | $\log_2 M = 2$ | $2N = 2000$ | $\log_2 M = 2$ |
| Number of Retraining Data | 400 | 4 | 40 |
| Transmission Overhead (bit) | 800 (1×) | 8000 (10×) | 80 (0.1×) |
| Retraining Overhead (s) | 11.24 (1×) | 0.64 (0.057×) | 2.67 (0.238×) |

Firstly, it can be seen that without the help of NN-CE, it suffers from dramatic transmission overhead and retraining overhead. Secondly, when $\mathbf{s}_{golden}$ is adopted as the label, retraining overhead can be reduced by 94% due to the benefit from the light NN-CE model. However, the data length of $\mathbf{s}_{golden}$ induces severe transmission overhead. Finally, the mechanism of end-to-end optimization can successfully reduce dramatic transmission overhead and retraining overhead by 90% and 76%, respectively. Therefore, we can adjust between these two efficient mechanisms for online retraining based on different requirements.

### E. Analysis of Computational Complexity

To evaluate the computational complexity, the number of different operations required by different classifiers and their corresponding inference time on a CPU are calculated and listed in Table V. In this part, we also consider the operations of transformation. The analysis is based on a signal with length $N$ among $M$ potential modulation candidates. $M_i$ denotes the alphabet size of the $i$th modulation candidate; $N_a = |\alpha|$ and $N_\theta = |\Theta|$ represent the size of the set of unknown parameters for HLRT.

From Table V, we can observe that cumulant [21] has the lowest complexity and inference time, but with the worst performance. Though ML [13] has optimal performance under AWGN channel, it is sensitive to channel mismatch and not practical for implementation. Therefore, DL-based approaches exploit the tradeoff between LB and FB approaches at the cost of memory usage. Besides, under the fading channel, the overhead of NN-CE is small enough with only a small increase in inference time, which is far less than HLRT [15], by 2200 times with significant improvement. Another interesting observation is that although our proposed approach has a smaller model size, it has a higher inference time than IQ-based [31]. This is due to the extra operations of the exponential function in Algorithm 3.

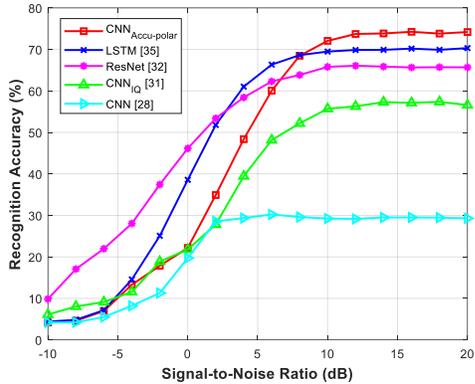

Fig. 12. Recognition accuracy against various SNR on RadioML dataset.

TABLE VI
RADIOML2018.01A DATASET PARAMETERS

| | |
|---|---|
| Modulation Types | OOK, 4ASK, 8ASK, BPSK, QPSK, 8PSK, 16PSK, 32PSK, 16APSK, 32APSK, 64APSK, 128APSK, 16QAM, 32QAM, 64QAM, 128QAM, 256QAM, AM-SSB-WC, AM-SSB-SC, AM-DSB-WC, AM-DSB-SC, FM, GMSK, OQPSK |
| Sample Length | 1024 |
| SNR Range | -10, -8, -6, -4, -2, 0, 2, 4, 6, 8, 10, 12, 14, 16, 18, 20 |
| Number of Training Data | 259,200 |
| Number of Validation Data | 28,800 |
| Number of Testing Data | 288,000 |

TABLE VII
COMPARISON OF MODEL SIZE AND TRAINING TIME BETWEEN PROPOSED METHOD AND PRIOR ARTS

| Models | Proposed Approach | IQ [31] | CNN [28] | ResNet [32] | LSTM [35] |
|---|---|---|---|---|---|
| Accuracy @ 20 dB | 74.2 | 56.6 | 29.3 | 65.7 | 70.3 |
| Model Parameters | 14,428 (1×) | 14,428 (1×) | 1,319,200 (91.4×) | 238,840 (16.6×) | 5,442,900 (377.2×) |
| Time per Epoch (s) | 6 | 6 | 18 | 46 | 347 |
| Number of Epoch | 33 | 42 | 35 | 40 | 32 |
| Total Training Time (s) | 198 (1×) | 252 (1.27×) | 630 (3.18×) | 1,840 (9.29×) | 11,104 (56.08×) |

Besides, the computational complexity of LB approaches increases rapidly with the number of potential modulation candidates $M$, in the scale of $\mathcal{O}(N)$ and $\mathcal{O}(N_a N_\theta N)$ which are not scalable. However, the proposed method is in the scale of $\mathcal{O}(1)$, meaning that we can support more modulation candidates without growth in computational complexity.

### F. Performance Evaluation on RadioML Dataset [47]

Finally, we evaluate the proposed accumulated polar-based approach on a publicly available dataset as a benchmark for a fair comparison. Besides, we also compare our approaches with prior arts of neural network-based classifiers, including one-dimensional CNN [28], IQ-based CNN [31], residual network (ResNet) [32], and LSTM [35]. Compared to earlier released RadioML2016.10a, the used RadioML2018.01a dataset in this work is much more difficult, which has 24 different modulations and was firstly evaluated in [32]. The dataset is synthetically generated using GNU Radio with a number of realistic channel imperfections, such as multipath fading, channel frequency offset, and sample rate offset [47].

In this dataset, it contains 24 digital and analog modulation types with SNR ranging from -20 dB to 30 dB and the space is 2 dB. However, due to the poor performance below -10 dB and no improvement over 20 dB, we only utilize the data with SNR between -10 dB and 20 dB. For each modulation type at each SNR value, we use 1500 examples with a length of 1024, which are separated into training, validation and testing dataset. The parameters of dataset are summarized in Table VI. For more details about this dataset, please refer to [32] and [47].

Firstly, we compare our proposed approach to prior arts under various SNR. Notice that the part of NN-CE is not included in this experiment and the model size $t$ for both IQ-based CNN [31] and the proposed accumulated polar feature is set to 4 due to this dataset is more difficult than our previous simulation settings. For the other prior arts, we implement the classifiers based on the proposed model architecture. In Fig. 12, we can observe that ResNet [32] can achieve the best performance in low SNR range and LSTM [35] becomes the best in middle range. On the other hand, our proposed approach has more than 4% improvement in high SNR range, which benefits from the proposed accumulated polar feature. We surmise that the degraded performance in low SNR range is due to the projection to grid-like image in our method may further distort the low-quality signals and thus results in unrecognizable patterns.

Furthermore, we compare the model size and training time between these approaches as shown in Table VII. From Table VII, we can observe that LSTM consumes about 377 times memory overhead and 56 times training time compared to our approach, which is intolerable for real applications. On the other hand, though ResNet can avoid significant memory overhead via max-pooling layer, its computational complexity is still too high to implement in communication systems. In conclusion, compared to state-of-the-arts, our proposed approach can transform the received data to an easier classified domain and thus achieve the best efficiency and lightness by dramatically reducing memory overhead and computational complexity.

Finally, the confusion matrix of the proposed method is shown in Fig. 13 for the analysis of classification accuracy between different modulation types. We can observe that there is serious misclassification appearing between high-order modulation types and AM modes, such as PSK, QAM, with-carrier (WC), and suppressed-carrier (SC), due to the similar symbol structure as concluded in [32].

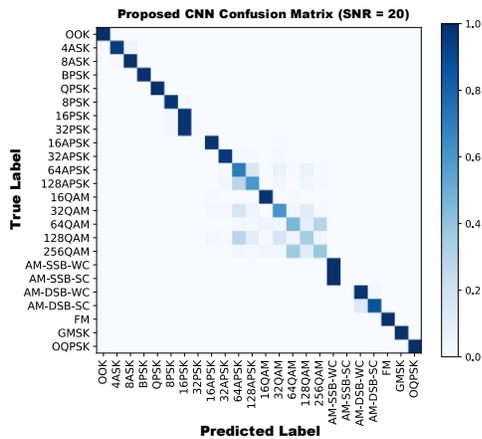

Fig. 13. Confusion matrix of the proposed accumulated polar feature on RadioML dataset at 20 dB.

In summary, our proposed accumulated polar feature has best recognition accuracy in high SNR range on this dataset with the following observations:
1. Benefiting from the domain knowledge for data preprocessing, the model size and computational complexity are far smaller, which is more feasible for practical implementation and resource-limited scenarios.
2. For further improvement of recognition accuracy, we provide two directions as future works. 1) The NN-CE can be utilized with dedicated design to compensate for the channel impairments, which makes the signal easier classified and thus avoids complicated and powerful neural networks for classification. 2) Inspired by [36], the proposed approach can be used in the first stage for fast and efficient classification and the indistinguishable modulation types, such as high-order PSK and AM-SSB in Fig. 13, can be further classified in the second stage with a more powerful classifier, which can improve both accuracy and system efficiency.

## VI. CONCLUSION

In this paper, a novel accumulated polar feature based deep learning with channel compensation mechanism for AMC is presented. Firstly, with the accumulated polar feature, it can learn from $r$-$\theta$ domain with historical information to reduce the offline training overhead and approach optimal recognition accuracy. Secondly, the proposed NN-CE can compensate for the distorted signal in realistic channel. Moreover, we propose two mechanisms for online retraining to deal with the time-varying fading channel; while having lower transmission overhead and retraining overhead. Therefore, the proposed design can be used as an efficient and lightweight technique for realizing intelligent receiver in practical environments, such as resource-limited IoT and V2X applications.

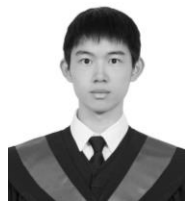
**Chieh-Fang Teng** (S'17) received his B.S. degree in electrical engineering from National Taiwan University, Taipei, Taiwan, in 2017. He is currently pursuing a Ph.D. degree in the Graduate Institute of Electronics Engineering, National Taiwan University. His research interests are in the areas of Internet-of-things, VLSI architecture for DSP, and 5G wireless communication technologies with the assisted of machine learning.

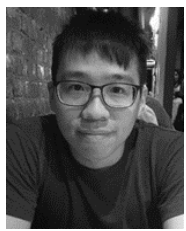
**Ching-Yao Chou** (S'16) received the B.S. degree in electrical engineering from National Taiwan University, Taiwan, in 2014. He is currently pursuing the Ph.D. degree in the Graduate Institute of Electronics Engineering, National Taiwan University. His research focuses on low-complexity signal processing for edge analysis, trying to link compressive sensing with machine learning.

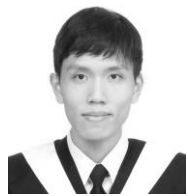
**Chun-Hsiang Chen** received the B.S. degree in electrical engineering from National Central University, Taoyuan, Taiwan, in 2018. He is currently pursuing the M.S. degree with the Graduate Institute of Electronics Engineering, National Taiwan University, Taipei, Taiwan. His research interests include communication systems and machine learning.

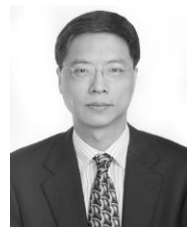
**An-Yeu (Andy) Wu** (M'96-SM'12-F'15) received the B.S. degree from National Taiwan University in 1987, and the M.S. and Ph.D. degrees from the University of Maryland, College Park in 1992 and 1995, respectively, all in Electrical Engineering. In August 2000, he joined the faculty of the Department of Electrical Engineering and the Graduate Institute of Electronics Engineering, National Taiwan University, where he is currently a distinguished professor. His research interests include VLSI architectures for signal processing and communications, and adaptive/multirate signal processing. He has published more than 190 refereed journal and conference papers in above research areas, together with five book chapters and 16 granted US patents. From August 2007 to Dec. 2009, he was on leave from NTU and served as the Deputy General Director of SoC Technology Center (STC), Industrial Technology Research Institute (ITRI), Hsinchu, Taiwan. In 2010, he received "Outstanding EE Professor Award" from The Chinese Institute of Electrical Engineering (CIEE), Taiwan. From 2012 to 2014, he served as the Chair of VLSI Systems and Applications (VSA) Technical Committee (TC), one of the largest TCs in IEEE Circuits and Systems (CAS) Society. In 2015, he is elevated to IEEE Fellow for his contributions to DSP algorithms and VLSI designs for communication IC/SoC.